\documentclass[twocolumn, showpacs,amsmath,amssymb,superscriptaddress]{revtex4}
\usepackage{graphicx}% Include figure files
\usepackage{dcolumn}% Align table columns on decimal point
\usepackage{bm}% bold math
\newcommand{\nc}{\newcommand}
\nc{\be}{\begin{equation}}
\nc{\ee}{\end{equation}}
\nc{\bea}{\begin{eqnarray}}
\nc{\eea}{\end{eqnarray}}
\nc{\bean}{\begin{eqnarray*}}
\nc{\eean}{\end{eqnarray*}}
\nc{\mb}{\mbox}
\nc{\rnc}{\renewcommand}
\nc{\vk}{\mb{\bf k}}
\nc{\vp}{\mb{\bf p}}
\nc{\vq}{\mb{\bf q}}
\nc{\vz}{\hat {\mb{\bf z}}}
\nc{\vj}{\mb{\boldmath$j$}}
\nc{\x}{\mb{\boldmath$x$}}
\nc{\A}{\mb{\boldmath$A$}}
\nc{\sa}{\mb{\boldmath$a$}}
\nc{\vs}{\mb{\boldmath$\sigma$}}
\nc{\nab}{\nabla}
\nc{\X}{\sf x}

\begin{document}

\title{Non-vanishing spin Hall currents in disordered spin-orbit coupling
systems}
\author{K. Nomura} 
\affiliation{Department of Physics, University of Texas at Austin,
Austin TX 78712-1081, USA}
\author{Jairo Sinova}
\affiliation{Department of Physics, Texas A\&M University, College
Station, TX 77843-4242, USA}
\author{T. Jungwirth}
\affiliation{Institute of Physics, ASCR, Cukrovarnick\'a 10, 162 53
Praha 6, Czech Republic}
\affiliation{School of Physics and Astronomy, University of Nottingham,
University Park, Nottingham NG7 2RD, UK} 
\affiliation{Department of Physics, University of Texas at Austin,
Austin TX 78712-1081, USA}
\author{Q. Niu}
\affiliation{Department of Physics, University of Texas at Austin,
Austin TX 78712-1081, USA}
\author{A.H. MacDonald}
\affiliation{Department of Physics, University of Texas at Austin,
Austin TX 78712-1081, USA}

\date{\today}

\begin{abstract}
Spin-orbit coupling induced spin Hall currents are 
generic in metals and doped semiconductors.
It has recently been argued that the spin Hall conductivity can be dominated by an intrinsic 
contribution that follows from Bloch state distortion in the presence of 
an electric field.  Here we report on an numerical demonstration of
the robustness of this effect in the presence of disorder scattering for the 
case of a two-dimensional electron-gas with Rashba spin-orbit interactions. 
\end{abstract}


\maketitle
  
Semiconductor spintronics research over the past decade has concentrated on 
the properties of spin-polarized carriers created by optical orientation, 
on the search for new ferromagnetic semiconductors with more favorable properties, and on
the injection of spin-polarized carriers into semiconductors from
ferromagnetic metals \cite{wolf,awschalom,dassarma}.
There has recently been a flurry of theoretical
interest \cite{dyakonov,hirsch,zhang,murakami,sinova,culcer,murakami2,schliemann,sinitsyn,shen,
inoue,burkov,rashba,murakami3,schliemann2,halperin,haldane,khaetskii,schwab,rashba2}
in the spin Hall effect \cite{dyakonov,hirsch,zhang}, {\it i.e.} in
transverse spin currents induced 
%in any metal or doped semiconductor
by an electric field.  Murakami \cite{murakami} {\it et al.} and Sinova \cite{sinova} {\it et al.} have 
argued in different contexts that the spin Hall conductivity can be dominated
by a contribution that follows from the distortion of Bloch electrons 
by an electric field and therefore approaches an intrinsic value in the 
clean limit.  This conclusion has recently been questioned, for the case of 
two-dimensional electrons with Rashba spin-orbit interactions in particular, by 
several authors\cite{inoue,rashba,halperin,khaetskii,schwab,rashba2} motivated by 
a number of different considerations, some of which are related to
controversies \cite{k-l,smit,berger,niu,jungwirth,fang,yao} that have 
long surrounded the theory of the
anomalous Hall effect in ferromagnetic metals and semiconductors.
In this Rapid Communication we report on a study based on numerically 
exact evaluation of the linear-response-theory  
Kubo-formula expression for the spin Hall conductivity.  We demonstrate that 
the intrinsic spin Hall effect is robust in the presence of disorder, 
falling to zero only when the life-time broadening 
energy is larger than the spin-orbit splitting of the bands.
The correlations between spin-orientation and velocity in the presence of an electric field that 
lie behind the intrinsic spin Hall effect {\em are not} diminished by weak
disorder.

We consider a two-dimensional electron system with the Rashba spin-orbit
interaction(R2DES):
\bea
        H = {\vp^2}/{2m}+\lambda [\vp\times\vz]\cdot\vs/\hbar\ +\ V.
\label{ham}
\eea
where $\vs$ is the Pauli matrix, $m$ is the effective mass, and $\lambda$ is the Rashba
spin-orbit coupling constant.
When the disorder potential $V$ in Eq.~(\ref{ham}) is absent, $\vp = \hbar \vk$ is a good
quantum number.  The Rashba spin-orbit interaction term can be viewed as Zeeman
coupling to a $\vk$-dependent effective magnetic field
${\bf {\Delta}}=(2\lambda)\vz\times\vk$.  The $V=0$ eigenstates are therefore 
the $S=1/2$ spinors oriented parallel and antiparallel to these fields:
$
|\vk \pm\rangle= 
\left[
\begin{array}{rr}
\mp ie^{-i\phi},
\    1
\end{array}
\right] e^{i\vk \cdot {\mb{\bf r}}}/\sqrt{2\Omega},
$
%Question1: The spinor in the previous draft was raised to the undefined 
% power t.  I have removed this - assuming that it is a type.  Am I missing something? 
and the two eigenvalues at a given $\vk$ are split by $2\lambda |\vk|$.   
Here $\phi=\tan^{-1}(k_x/k_y)$, $\Omega$ is the system area 
and we have applied periodic boundary conditions.
As explained in Ref.\cite{sinova}, an electric field in the
$x$-direction causes Rashba spinors to tilt out of the $x$-$y$ plane
giving rise to an intrinsic spin Hall effect.
The key issue in dispute is whether or not
the velocity-dependent spinor tilts vanish when quasiparticle 
disorder scattering is properly taken
into account.  To address this subtle issue without making any 
assumptions which might prejudice the conclusion,
we evaluate the Kubo formula for the spin Hall conductivity using the exact 
single-particle eigenstates of a disordered 
finite area two-dimensional electron system with Rashba
spin-orbit interactions. 

Our disorder potential consists of randomly centered scatterers
that have strength $u_0$ and a Gaussian spatial profile with range
$l_v$.  
The potential matrix elements satisfy
$\overline{|\langle \vk \sigma|V|\vk'\sigma'\rangle|^2} = (n_iu_0^2/\Omega)\delta_{\sigma\sigma'}
\exp(-|\vk-\vk'|^2l_v^2)$, where the density of scatterers $n_i$ (intended to represent
remote ionized donors) is set equal to the electron density.
%Question2:  I think that we should make a more specific statement than `comparable'
% What did we actually do.  I changed `comparable' to equal above.  If this is not correct
% it should be replaced by the correct statement.
It is widely recognized that 2DES disorder potentials can have long correlation lengths 
up to $\sim 100$ [nm]. To examine how our conclusions depend on the range of the disorder potential, we 
have performed calculations for correlation lengths ranging from $l_v \sim 0$ to
$l_v \sim 100$[nm].

We diagonalize the Hamiltonian in the $\lambda=0$ eigenstate representation and 
introduce a hard cutoff at a sufficiently large momentum $\Lambda$.
For a fixed particle density, the number of electron $N_e$ and 
the system size are related by $\Omega=L^2=N_e/n_e$.  Our conclusions are based 
on calculations with $N_e$ up to $2258$.
For $n_e=0.6\times 10^{11}$ [cm$^{-2}$] the system size is up to
$L=2\ [{\mathrm \mu }$m], longer than the characteristic microscopic
length scales, the mean-free path ($l\sim 10^2-10^3$ [nm]),
the Fermi wavelength ($\lambda_F=2\pi/k_F=101$ [nm]), and the disorder potential
range ($l_v \le100$ [nm]). The system size in these simulations 
is comparable to that of typical 2DES channels in electronic devices.  
We fix the effective mass at the bulk GaAs value, $m=0.067m_{e}$,
where $m_e$ is the bare electron mass and perform calculations over a wide
range of $\lambda$ and $u_0$ values.

The Kubo formula expression for the $z$ spin component of the spin Hall conductivity is:
\be
\sigma_{\mu\nu}^z(\omega)=\frac{1}{i\Omega}\sum_{n,n'}
\frac{f(E_n)-f(E_{n'})}{E_n-E_{n'}} \; 
\frac{\langle n|j_{\mu}^z|n'\rangle\langle
n'|j_{\nu}|n\rangle}{\hbar\omega+E_n-E_{n'}+i\eta},
\label{kubof}
\ee
where $f(E)$ is the Fermi function, $n$ labels 
exact eigenstates with eigenvalues $E_n$,
% of the $n$-th single-particle state, 
and the charge and spin current operators are 
$
{\vj}=-e\ \partial H/\partial {\bf p}=-e \left({\vp}/{m}+\lambda
{\vz}\times\vs/\hbar\right)
$ and
${\vj}^{z}=\{\partial H/\partial {\bf p},\frac{\hbar}{2}\sigma_{z}\}/2 = {\bf p}\, \sigma_{z}/m 
$ respectively \cite{sinova}.  In finite size calculations the electric field turn on time
$\eta^{-1}$  
must be shorter than the transit time in the simulation cell in order to 
obtain the correct thermodynamic limit for the conductivity.  In the metallic
limit of interest here, $\eta$ must exceed the simulation cell level
spacing but be smaller than all intensive energy scales.  
In the dc $\omega=0$ limit, $\sigma_{\mu\nu}^z$ is real with a dissipative contribution that comes 
from the $i\eta$ term in the denominator and a reactive contribution that 
comes from the imaginary part of the matrix element product.
%The latter contribution reduces to the intrinsic spin Hall conductivity in the %absence of disorder.

Typical numerical results for the disorder and spin-orbit coupling 
strength dependence of the spin Hall conductivity
$\sigma_{sH}=\sigma_{xy}^z(\omega=0)$ are illustrated in Fig.1.  (These  
calculations are for $l_v\sim 80$ [nm].)
We find that in the strong Rashba coupling, weak-disorder regime  %($\epsilon_F\tau \stackrel{>}{_\sim}  20$)
the spin Hall conductivity is close to the (universal) intrinsic value for this model,
and that it decreases for weaker spin-orbit coupling and stronger
disorder. Experimentally, Rashba spin-orbit coupling strength can be varied 
over a wide range by tuning a gate field \cite{nitta,koga}.
We have varied the spin-orbit coupling strength at the Fermi energy $\lambda
k_F$ from $0.1 \epsilon_F$ to $0.4 \epsilon_F$. The system size for the calculations summarized by 
Fig.1 was $1500 {\rm nm}$.  The range we have chosen for disorder strength
values was based on the golden-rule expression for the transport scattering 
rate\cite{mahan},
$\hbar/\tau = {2\pi}\sum_{\vk'}|V(\vk-\vk')|^2(1-{\hat \vk} \cdot {\hat \vk'})\delta(\epsilon_{\vk'}-\epsilon_F).
$
%Here $\tau$ has a unit [meV$^{-1}$].
The golden-rule combined with Boltzmann transport theory yields the Drude expression for the
longitudinal conductivity,
$\sigma_D=ne^2\tau/m=2\epsilon_F\tau(e^2/h).$
Using these approximate estimates, we have varied the disorder strength so that $\epsilon_F\tau$ 
covers the range $2-20$, typical for two-dimensional electron systems.
%Question3:  The words that I have chosen are based on the assumption that $l_v$ was fixed for
%Fig.1 only.  If this is incorrect, the words should be changed. 
For GaAs materials parameters, the disorder strength range that we consider corresponds
to mean-free paths $l=70-700$ [nm].
We note that in the case of short-range scatterers ($l_v\sim 10$ [nm]) the transport lifetime $\tau$ defined above is not so
different from the momentum lifetime $\tau_0$ given by $\hbar/\tau_0=
{2\pi}\sum_{\vk'}|V(\vk-\vk')|^2\delta(\epsilon_{\vk'}-\epsilon_F)$ ($l_v\sim 10$ [nm]), whereas these 
quantities differ substantially for longer (and more realistic) 
correlation lengths.  In what follows we take $\hbar =1$ so that $\tau^{-1}$ has energy units. 
\begin{figure}[!t]
\includegraphics[width=0.45\textwidth]{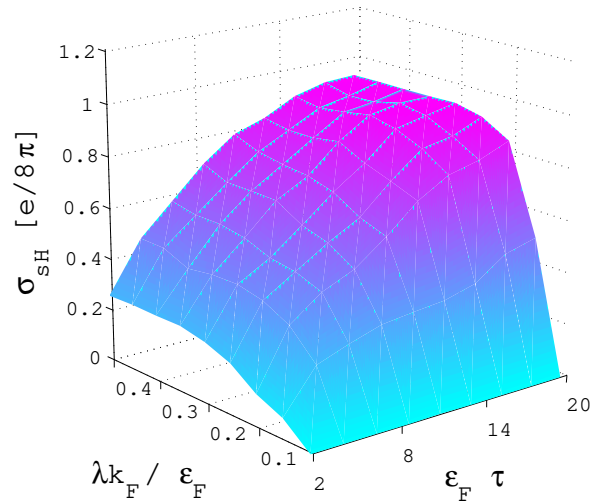}
\caption{
Spin Hall conductivity $\sigma_{ sH}^{}$ as a function of $\epsilon_F\tau$ and $\lambda
 k_F /\epsilon_F$ at $\epsilon_F=2.15$ [meV] and $n_e=0.6\times10^{11}$
 [cm$^{-2}$]. For these calculations the system size is $L=1500$ [nm] and $l_v=80$ [nm].
Note that the conductivity depends mainly on $\lambda k_F \tau$ and that,
because our interest is limited to the metallic regime, our calculation range 
does not address the strong scattering limit $\tau \to 0$. 
}
\label{}
\end{figure}
These results demonstrate that for this model
$\sigma_{sH}$ is to reasonable accuracy a function of only $\lambda k_F\tau$,
the ratio of the spin-orbit splitting to the quasiparticle state lifetime broadening.
%and roughly consistent with disorder treatments that neglect\cite{sinova,schliemann,sinitsyn}
%that neglec
%\cite{sinova,schliemann,sinitsyn} even rather than taking into
%account the vertex corrections from ladder diagrams \cite{inoue}.
The intrinsic spin Hall conductivity survives provided that 
$\lambda k_F \tau > 1.$

\begin{figure}[!b]
\begin{center}
\includegraphics[width=0.48\textwidth]{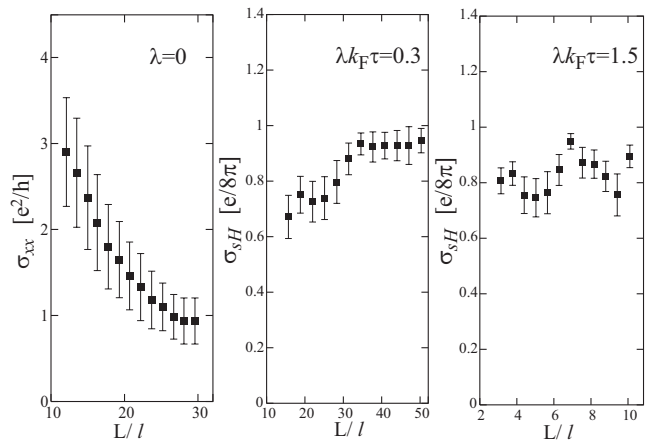}
\caption{Left: Size dependence of the longitudinal conductivity $\sigma_{xx}$
 as a function of $L/l$ for $\lambda=0$,
$\epsilon_F \tau=2.0$, and $l_v=20$ [nm];
Middle and Right: $L/l$ dependence of the spin Hall conductivity $\sigma_{SH}$ for
$l_v=20$ [nm]) and $\lambda k_F/\epsilon_F=0.3$.
The middle panel is for a strongly disordered system in which $\epsilon_F \tau =1$
while the right panel is for the a weakly disordered system in which $\epsilon_F \tau =5$.
}
\label{fig:two}
\end{center}
\end{figure}

Fig. 2 illustrates some typical system size dependences of the finite-size 
longitudinal $\sigma_{xx}$ and spin Hall $\sigma_{sH}$ conductivities. 
The size-dependence of transport coefficients in disordered systems 
can reflect quantum corrections to Boltzmann
transport theory due to the interference effects that cause localization.
In two-dimensions, scaling theory and microscopic perturbative
calculations predict $\sigma_{xx}$ corrections that depend on
spin-orbit coupling strength and can grow when 
the system size $L$ is larger than the mean-free path $l$.  The 
conductivity is expected to decay
exponentially with system size in the strongly localized region.\cite{lee}
Numerical $\sigma_{xx}$ results for the strongly disordered 
case $\epsilon_F\tau=2$, $\lambda=0$, and $l_v=20$ [nm],
shown in the left panel of Fig.2, are consistent with expectations for this
thoroughly studied quantity.\cite{lee} 
%Question4: Should we add a reference above? 
Our main interest at present, however, is the system size dependence of
the spin Hall conductivity $\sigma_{ sH}$ and particularly in establishing
whether or not it vanishes in the limit $L \to \infty$.  For 
$\sigma_{ sH}$, $L$ should be compared with both $l$ and with the spin-orbit
length $L_{so}= l/(\lambda k_F \tau)$.
In the middle panel of Fig.[~\ref{fig:two}] $L_{so} \approx 3 l$ is the 
longer intensive length scale, with some system size apparent up to $L/L_{so} \sim 10$.
For the more weakly disordered case in the right panel 
$l$ is longer and no systematic $L/l$ dependence was found. 
These numerical results appear to establish rather unambiguously that 
$\lim_{L \to \infty}\sigma_{ sH} \ne 0$.  
%  Note that for the most
%weakly disordered cases illustrated, the results in the right panel of Fig.2 are close
%to the ballistic limit and therefore do not really shed light on the large $L/l$ limit.

The intrinsic spin Hall effect in the R2DEG is due to a correlation
\cite{sinova} between quasiparticle velocity and the $z$-component
of spin induced by an electric field; for an electric field in the
$x$-direction, an up spin is induced in positive 
$y$-component velocity majority-band states and a corresponding down spin at negative 
velocities. After summing over bands, 
coherence is confined in momentum space to the annulus of singly-occupied states.  
These responses are induced by the  
interband matrix elements of the perturbation term in the Hamiltonian 
that accounts for the spatially uniform electric field.
Since the observable we are interested in here, the spin Hall current,
is purely off-diagonal in band indices, its response depends on 
interband coherence alone and not at all on the altered Bloch 
state occupation probabilities that dominate most transport coefficients
in metals and are the focus of Boltzmann transport theory.
If the spin Hall conductivity were to vanish because of disorder 
scattering, the intrinsic interband coherence would either have to be
cancelled at all wavevectors, or be cancelled by stronger coherences induced in  
a narrow transport window (presumably of width $1/\tau$) centered 
on the Fermi circles. 

In Fig.3 we compare the exact linear-response 
momentum-dependent $z$-direction spin-density (and hence interband coherence)
for a disorder-free system (left panel) with $\lambda k_F/\epsilon_{F} = 0.2$ 
%Question: We need the value of \lambda k_F for this calculation.
with that of a disordered system (right panel) with the same spin-orbit
interaction strength and $ \epsilon_{F} \tau = 3.2$.  ( $l_v/\lambda_F=0.2$ for 
the calculations illustrated in Fig.3.) Both quantities
are proportional to the electric field and are plotted in the same units. 
These results were obtained from the same linear response theory expressions
used in Eq.(2) with
$S_z({\vk})=\sum_{\sigma}\, \sigma/2 \,|{\vk}\sigma\rangle\langle{\vk}\sigma|=
(|\vk +\rangle\langle\vk -|+|\vk -\rangle\langle\vk +|)/2$
substituted for the spin current $j^z_{\mu}$.
The disorder averaged spin Hall conductivity and longitudinal conductivity in this case are
$\sigma_{sH}/(e/8\pi)=0.64$ and $\sigma_{xx}/(e^2/h)=5.1$  at $\epsilon_F\tau=3.2$.
Our numerical calculations demonstrate that the coherence is not changed
qualitatively by impurity scattering, maintaining the same angle dependence
as it is spread in momentum space.  In particularly there is no evidence that the 
direction averaged coherence is either cancelled uniformly or cancelled by a 
strong contribution more narrowly centered on the two Fermi circles.  

\begin{figure}[!t]
%\begin{center}
\includegraphics[width=0.56\textwidth]{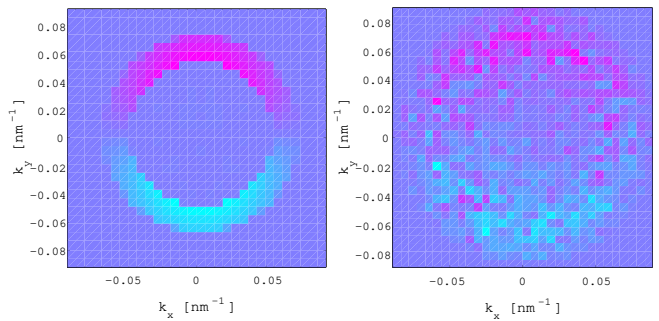}
\caption{
Electric field induced spin distribution $S_z(\vk)/eE$ as a function of
wave vector for the clean limit (left side) and for an $\epsilon_F\tau=3.2$
$l_v\lambda_F=0.2$ disorder model (right side).
}
\label{}
%\end{center}
\end{figure}

The subtleties that confuse theories of the spin Hall conductivity in 
a R2DES are related to issues 
that arise quite generally in the linear-response theory analysis of non-dissipative
transport coefficients, like the anomalous Hall conductivity \cite{noteonahe} of a ferromagnet,
the ordinary Hall conductivity of a paramagnet, and the spin Hall conductivity of 
other paramagnetic metals.  From an exact eigenstate Kubo formula point of view, these transport coefficients 
can be dominated by reactive contributions that come from states far from the
Fermi level and are not associated with electric field induced level crossings and dissipation.
In the spin and anomalous Hall effect cases, the reactive contributions
do not vanish in the limit of a perfect crystal, instead approaching an intrinsic value.
The currents accounted for by these intrinsic Hall coefficients can be viewed as corresponding 
to equilibrium currents that flow in an effective periodic systems whose symmetry has been reduced by the
electric field.  This point has been emphasized recently by Rashba\cite{rashba2},
who argues on this basis that the intrinsic response is 
a transient that will be attenuated within a relaxation time $\tau$ scale after the 
electric field is turned on.  Similar arguments have been made concerning the 
intrinsic contribution to the anomalous Hall effect.\cite{smit}  The specific instance 
studied here is perhaps an especially simple example of this class of effects, precisely because
$S_z(\vk)$ and the spin Hall current are purely off-diagonal in band indices.
We conjecture, as an extrapolation from the present numerical study, that 
the part of the density-matrix linear response that is off-diagonal in band index 
always approaches its intrinsic value in the weak disorder limit.  The spin Hall current
operator, like the charge current operator in the case of the anomalous Hall effect,
will also  have intraband matrix elements in the general case.  We expect that
these can in general lead to extrinsic intraband contributions to the linear response conductivity that 
remain finite in the weak disorder scattering limit.    

In a realistic sample with boundaries, spin density is accumulated at 
the sample edge by the spin currents. We expect that edge spin accumulations can be measured experimentally.
Stevens {\it et al}.\cite{stevens} have recently 
reported on a remarkable optical measurement of accumulation due to non-linear response
spin currents using a spatially resolved pump-probe technique in GaAs/AlGaAs quantum
wells. Similar luminescence polarization measurements should be able to 
detect electrically generated linear response spin Hall currents. 

In summary, we calculated the spin Hall conductivity in a
disordered system with Rashba spin-orbit coupling using the exactly
evaluated eigenstates of the Hamiltonian and the Kubo linear response theory. 
We find that the field induced spin Hall current of this model
approaches its intrinsic value in the limit of weak disorder 
scattering. 

The authors thank G. Bauer, D. Culcer, E. M. Hankiewicz, J. Inoue, L. Molenkamp, 
S. Murakami, E. Sherman, N.A. Sinitsyn, X.C. Xie, and
S.-C. Zhang for useful discussions.  One of the authors K.N. is supported by 
the Japan Society for the Promotion of Science by a  
Research Fellowship for Young Scientists.  This work has been supported by 
the Welch Foundation and by the Department of Energy under grant DE-FG03-02ER45958.

{\it Note added}. ---After this work was completed and submitted several preprints
appeared reporting on related numerical simulations\cite{nikolic,sheng,hankiewicz}
of spin Hall conductance in finite samples with contacts.  These studies reach
similar conclusions on the robustness of spin Hall effects. 
Very recently two experimental preprints\cite{awschalomscience,wunderkind} have appeared which report detection 
of edge spin accumulation due to spin Hall currents.
% Produces the bibliography via BibTeX.

\end{document}